\documentclass[10pt,twocolumn,letterpaper]{article}

\usepackage{geometry}
\usepackage{times}
\usepackage{epsfig}
\usepackage{graphicx}
\usepackage{amsmath}
\usepackage{amssymb}
\usepackage{bbm}
\graphicspath{{images/}}
\DeclareMathOperator*{\argmin}{arg\,min}
\newcommand{\MM}{\cal{M}}
\newcommand{\SSS}{\cal{S}}

\usepackage[pagebackref=true,breaklinks=true,letterpaper=true,colorlinks,bookmarks=false]{hyperref}

\begin{document}

\title{Geodesic Distance Descriptors}
\author{Gil Shamai \and  Ron Kimmel
\thanks{Computer Science Department, 
Technion, Israel Institute of Technology, Haifa 32000.
}
}



\maketitle

\begin{abstract}
The Gromov-Hausdorff (GH) distance is traditionally used for measuring distances
 between metric spaces. 
It was adapted for non-rigid shape comparison and matching of isometric surfaces, 
 and is defined as the minimal distortion of embedding one surface into the other, 
  while the optimal correspondence can be described as the map that minimizes this distortion. 
Solving such a minimization is a hard combinatorial problem that requires pre-computation
 and storing of all pairwise geodesic distances for the matched surfaces. 
A popular way for compact representation of functions on surfaces is by projecting them 
 into the leading eigenfunctions of the Laplace-Beltrami Operator (LBO). 
When truncated, The basis of the LBO is known to be the optimal for representing functions
 with bounded gradient in a min-max sense. 
Methods such as Spectral-GMDS exploit this idea to simplify and efficiently approximate 
 a minimization related to the GH distance by operating in the truncated spectral domain, 
 and obtain state of the art results for matching of nearly isometric shapes. 
However, when considering only a specific set of functions on the surface, such as 
 geodesic distances, an optimized basis could be considered as an even better alternative.
Moreover, current simplifications of approximating the GH distance 
 introduce errors due to low rank approximations and relaxations of the permutation matrices.

Here, we define the geodesic distance basis, which is optimal for compact approximation 
 of geodesic distances, in terms of Frobenius norm. 
We use the suggested basis to extract the Geodesic Distance Descriptor (GDD), which encodes
 the geodesic distances information as a linear combination of the basis functions. 
We then show how these ideas can be used to efficiently and accurately approximate the 
 metric spaces matching problem  
 with almost no loss of information. 
We incorporate recent methods for efficient approximation of the proposed basis and 
 descriptor without actually computing and storing all geodesic distances. 
These observations are used to construct a very simple and efficient procedure 
 for shape correspondence. 
Experimental results show that the GDD improves both accuracy and efficiency of 
 state of the art shape matching procedures.
\end{abstract}
\section{Introduction}
One line of thought in shape analysis considers an object as 
 a metric space, and object matching, classification, and comparison as the 
 operation of measuring the discrepancies and similarities between such metric spaces,
 see, for example,
 \cite{elad2003bending}, \cite{zigelman2002texture}, \cite{schwartz1989numerical}, \cite{memoli2007use}, \cite{bronstein2006generalized}, \cite{Aflalo2016}, \cite{memoli2005theoretical}.
 
Although theoretically appealing, the computation of distances between metric spaces poses 
 complexity challenges as far as direct computation and memory requirements are involved. 
As a remedy, alternative representation spaces were proposed
\cite{ovsjanikov2012functional}, \cite{lipman2009mobius}, \cite{gu2004genus}, \cite{chen2015robust}, \cite{shtern2015spectral}, \cite{lai2014multi}, \cite{ling2007shape}.
The question of which representation to use in order to best represent the metric 
 space that define each form we deal with, and yet allow for an accurate representation 
 of the mapping from one metric space to another occupied the attention of some recent 
 efforts, see for example 
 \cite{aflalo2015optimality} and \cite{aflalo2016best}.
Indeed, some compact spaces, in the case of matching metric spaces, allow to 
 reduce the complexity and consequently improve the accuracy of the resulting
 correspondence maps between surfaces. 

As a specific example, both
 {\em Spectral generalized multidimensional scaling} (SGMDS) \cite{Aflalo2016} and functional maps
 \cite{ovsjanikov2012functional} try to find a linear mapping in a dual space that encodes the
 minimal distance distortion mapping between two shapes. 
By trying to match geodesic distances, the SGMDS efficient procedure provides accurate 
 correspondence maps between nearly isometric shapes.
Nevertheless, casting the Gromov-Hausdorff related minimization as is, into the spectral domain forces 
 a low rank representation of a relaxed version of the permutation matrix that encodes the
 correspondence, and introduces errors.

When considering the set of all gradient bounded functions on a given manifold, it can be shown
 that the eigenfunctions of the Laplace-Beltrami operator (LBO) provide an optimal and unique,
 in a min-max sense, representation for truncated bases \cite{aflalo2015optimality}.
Although not explicitly acknowledged at the time, it motivated its usage in shape matching methods
 like the SGMDS and functional maps.
When dealing with a smaller subset of these functions, such as geodesic distances, 
 there could be a basis that would provide a better representation.
We introduce the geodesic distance basis for optimal representation of geodesic distances. 
We then construct the Geodesic Distance Descriptor (GDD) which encodes the geodesic distances 
 information as a linear combination of the basis functions.
The GDD can be seen as a canonical form without the metrication error.
It is shown that an approximated metric space matching minimization can be reduced to comparing the GDD 
 of the shapes using iterative closest point (ICP) procedures
 \cite{besl1992method}, \cite{chen1992object}, without truncation or relaxation.
The result is an accurate correspondence permutation matrix.
The new linear formulation significantly improves the computational
 complexity  required to solve the shape matching problem.
Moreover, when casting the problem on the truncated geodesic distance basis, almost no information
 is lost.
As a stand-alone method, GDD outperforms both SGMDS and functional maps in accuracy, efficiency, 
 and simplicity.
When combined with any of the state of the art methods, superior results are obtained.

In Section \ref{sec:GH} we define the Gromov-Hausdorff distance and its use 
 for shape correspondence.
In Section \ref{sec:Basis} we define the optimal basis for geodesic distance representation, 
 and relate it to the LBO basis.
Next, Section \ref{sec:NMDS} deals with efficiently approximating the geodesic distance basis without actually computing all pairwise geodesic distances.
The induced geodesic distance descriptor is defined in Section \ref{sec:GDD}, where we show how
 it can be used to approximate the solution of the metric space matching minimization problem.
In Section \ref{sec:FGMDS} we discuss a few initialization and post-processing alternatives to
 our final shape correspondence procedure that provides state-of-the-art results and presented
 in Section \ref{sec:Results}.


\section{The Gromov-Hausdorff Distance}
\label{sec:GH}

Given two shapes ${\SSS}_1$ and ${\SSS}_2$, consider the map that best preserves the 
 inter-geodesic distances while embedding one shape into the other.
The {\em Gromov-Hausdorff} (GH) distance is defined as the distortion of that embedding.
Let, $d_1(s,s')$ and $d_2(q,q')$ represent the inter-geodesic distances between $s, s' \in {\SSS}_1$ 
 and $q, q' \in {\SSS}_2$, respectively.
The GH distance is defined as
\begin{equation}
d_{GH}({\SSS}_1, {\SSS}_2) = \frac{1}{2} \min_{\cal{C}}\max_{(s,q)\in {\cal{C}},(s',q')\in {\cal{C}}} \big|d_1(s,s')-d_2(q,q')\big|
\end{equation}
 where
\begin{equation}
 \forall s\in {\SSS}_1 \ \ \exists q \in {\SSS}_2 \text{  s.t.  } (s,q) \in \cal{C},
\end{equation}
 and
\begin{equation}
\forall q\in {\SSS}_2 \ \ \exists s \in {\SSS}_1 \text{  s.t.  } (s,q) \in \cal{C}.
\end{equation}
The set of corresponding points is represented by $\cal{C}$.
{\color{red} }
The set $\cal{C}$ could be defined through an indicator function $p(s,q)$ such that $p(s,q) = 1$ if $(s,q) \in \cal{C}$ 
 and $p(s,q) = 0$ for $(s,q) \notin \cal{C}$. 
In practice, we detect correspondences between well sampled manifolds, for which we can re-write
 our optimization problem in matrix notation that reads
\begin{equation}
\label{eq:GH_inf}
\argmin_{P \in \pi(n)} \|PD_1P^T-D_2\|_\infty,
\end{equation}
 where $\pi(n)$ is the set of $n \times n$ permutation matrices, and $P,D_1, D_2$ 
  are the discretizations of $p(s,q), d_1(s,s'), d_2(q,q')$.

Several variations were proposed to reduce the complexity of the problem 
\cite{bronstein2006generalized}, \cite{bronstein2008numerical}, 
\cite{lipman2009surface}, 
\cite{navazosparse}.
In the Generalized Multi-Dimensional Scaling (GMDS) \cite{bronstein2006generalized}, 
 the $L_{\infty}$ Hausdorff distance was replaced by an $L_2$ norm.

SGMDS \cite{Aflalo2016} further simplifies this minimization by relaxing the permutation matrix $P$ and 
 reformulating it as
\begin{eqnarray}
& & \argmin_P  \|PA_1D_1-D_2A_2P\|_{{\SSS}_1, {\SSS}_2} \cr
& & \text{s.t.} \cr
& &  \,\,\, PA_1 \mathbbm{1}=\mathbbm{1}\cr
& &  \,\,\, P^TA_2 \mathbbm{1}=\mathbbm{1} 
\end{eqnarray}
 where $\|F\|_{{\SSS}_1, {\SSS}_2} = \text{trace}(F^TA_2FA_1)$, 
 and $A_1, A_2$ hold the infinitesimal areas about each sample point of ${\SSS}_1, {\SSS}_2$
  along their diagonals. 
Then, in order to be able to solve this minimization in a practical computational complexity, 
 $P$, $D_1$ and $D_2$ are represented in the truncated spectral domain of the LBO, 
 such that only the first $k$ eigenfunctions are considered. 
The number of variables in the simplified SGMDS minimization is quadratic in the number of eigenfunctions,
 so the optimization becomes relatively slow when considering more then $k = 20-30$ eigenfunctions. 
In addition, using only $k$ eigenfunctions to represent the permutation matrix $P$ forces it to 
 be represented as a $k$-rank matrix, which introduces 
  significant errors to the minimization.

In the following sections of this paper, we will propose an alternative to \ref{eq:GH_inf} that does not require any relaxation or truncation of $P$, while allowing much more eigenfunctions to be incorporated.
In addition, we will work with a basis that is optimized for geodesic distance representation, for which truncating the eigenspace almost does not affect the solution. 

\section{Geodesic Distances Basis}
\label{sec:Basis}
The set of eigenfunctions of the Laplace-Beltrami operator of some manifold $\MM$ form a basis, 
which generalizes the Fourier basis to surfaces and is widely used for representation of functions on manifolds.
When considering the set of all gradient bounded functions on $\MM$, the LBO eigenfunctions is the best basis 
 for a truncated representation of this set in a min-max sense \cite{aflalo2015optimality}. 
These eigenfunctions were used for a truncated representation
 of descriptors, permutation functions \cite{ovsjanikov2012functional}, and geodesic distance functions \cite{Aflalo2016}.
However, when considering a subset of these functions, such as geodesic distances, an optimized basis 
 could provide an even better representation.
In this section, we define the {\em Geodesic Distance Basis} and show its usefulness for compactly 
 representing geodesic distance functions. 
For simplicity, we limit our discussion to the discrete domain with the standard inner product, assuming the 
 shapes were sampled uniformly. 
All definitions and derivations could be easily extended to the more general case of non-uniform sampling.

Assume we are given a shape $\SSS$ with $n$ vertices, sampled from a smooth manifold. 
Let the $n \times n$ symmetric matrix $D$ hold all geodesic distances of $\SSS$, 
 such that $D_{ij}$ holds the geodesic distance between $i,j \in {\SSS}$.
%
Denote by $D = Q\Lambda Q^T$ the eigenvalue decomposition of $D$, where the columns of $Q$ are orthonormal
 and $\Lambda$ is a real diagonal matrix.
Assume that the set of columns of $Q$ are ordered by the magnitude of their corresponding eigenvectors, 
 in a descending order.
The $k$-truncated eigenvalue decomposition of $D$ is defined by $\hat D_Q = Q_k\Lambda_k Q_k^T$, 
 where $\Lambda_k$ holds the $k$ first eigenvalues along its diagonal and $Q_k$ holds the 
 first $k$ corresponding eigenvectors.
In general, it is known that the best $k$-rank approximation of a matrix, in terms of Frobenius norm, 
 is given by computing its $k$-truncated singular value decomposition. 
For symmetric matrices, it is equivalent to the $k$-truncated eigenvalue decomposition.
This can be formulated as 
\begin{equation}
	\hat D_Q = \argmin_{\hat D \in \mathcal{K}(n)}\|D - \hat D\|_F,
\end{equation}
 where $\mathcal{K}(n)$ is the space of $n\times n$ matrices with rank $k$.

We term the set of columns of $Q$ as the basis of geodesic distances of the shape $\SSS$. 
Notice that 
 $\hat D_Q = Q_k Q_k^T D$.
Let the matrix $B_k$ hold $k$ vectors of some other basis.
The truncated representation of $D$ in the new basis is obtained by 
 $\hat D_B = B_k B_k^T D$, where $B_k^T D$ are the coefficients of representation. The rank of $\hat D_B$ is at most $k$ as a product of $n \times k$ matrices. 
Hence, it cannot approximate $D$ better than $\hat D_Q$.
In other words, the truncated reconstruction of $D$ using the geodesic distance basis has the lowest
 approximation error, in terms of Frobenius norm, among all other bases, independent of the number of vertices $n$.

\begin{figure}[htbp]
\begin{center}
\includegraphics[width=0.49\columnwidth]{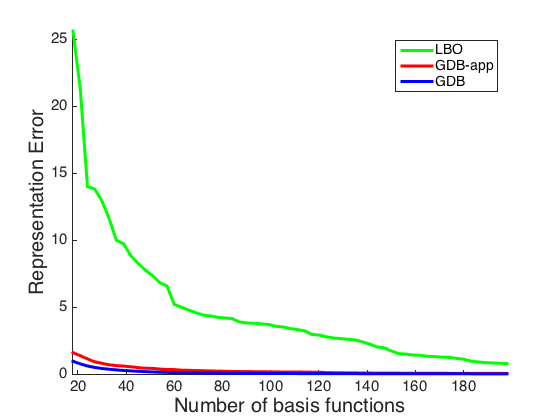}
\includegraphics[width=0.49\columnwidth]{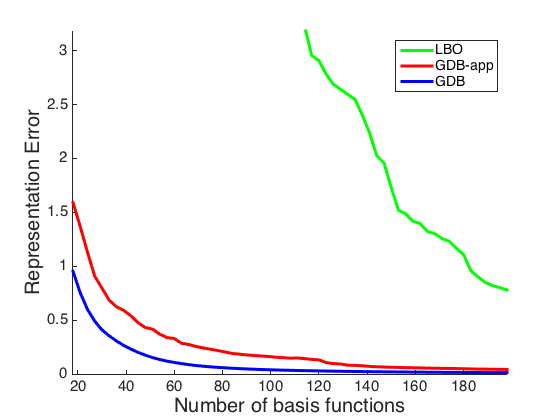}
\end{center}
\caption{Left - Comparing the reconstruction errors $\|D - \hat D\|_F$, where $\hat D$ is obtain using the basis of the LBO (LBO), the suggested optimal basis $Q$ (GDB), and its approximation $\tilde Q$ (GDB-app). The right image is a zoom of the left one.}
\label{fig:compare_basis}
\end{figure}

The computation of $Q$ is actually not practical when dealing with more than a few thousand points. However, it can be efficiently approximated.
In the next section we discuss on how to compute an approximation to $Q$.
In Figure \ref{fig:compare_basis} we compare the truncated reconstruction error of the geodesic distances obtained using the basis of the LBO to that of the proposed Geodesic Distance Basis. 
Here, we computed $Q$ on the Wolf shape from TOSCA, which is the only shape with less then ten thousand vertices.
It can be seen that the proposed basis supplies a compact representation with a much better reconstruction.
\section{Computing the basis}
\label{sec:NMDS}
In order to compute $Q$ and $\Lambda$, one would have to compute all pairwise geodesic distances in the large $n \times n$ matrix $D$, and then perform eigenvalue decomposition to obtain the largest magnitude $k$ eigenvalues and corresponding eigenvectors.
The task of computing all pairwise geodesic distances is time consuming and impractical when dealing with more than a few thousand points, even when using efficient methods such as \textit{Fast Marching} \cite{kimmel1998computing}.
In SGMDS, Spectral-MDS (SMDS) \cite{aflalo2013spectral} was adopted for an efficient computation of $D$. There, the geodesic distances were computed between roughly $2000$ samples of the shape, and the rest of the distances were interpolated by minimizing a derichlet energy term while working in the truncated spectral domain of the LBO. The geodesic distances were computed using fast marching and the samples were chosen using the \textit{Farthest Point Sampling} procedure \cite{hochbaum1985best}.

Recently, an alternative efficient implementation to Multidimentional Scaling (MDS) was suggested in \cite{shamai2015classical}. 
There, geodesic distances were interpolated from a few of them, similar to SMDS, and with the same complexity. However, no truncated representation in any basis was used, avoiding the errors caused by the truncation, and significantly increasing the accuracy of the approximation.
It was shown that only $p = 100$ samples were enough to reconstruct the geodesic distances up to negligible errors.
Eventually, the approximation to the pairwise geodesic distances matrix was written as a product of smaller matrices $S_{n \times k}T_{k \times k}S_{n \times k}^T$, where $k$ is half the size of the number of samples $p$.

Here, we adopt this idea to compute an approximation to the $k$ first basis functions of the geodesic distance basis.
Assume we have a decomposition $\hat D_{\tilde Q} = STS^T$, obtained from \cite{shamai2015classical}, that well approximates $D$, where $S$ is an $n \times k$ matrix and $T$ is a $k \times k$ matrix. 
denote by $\mathcal{Q}R$ the QR factorization of $S$, where $\mathcal{Q}$ is orthonormal and $R$ is upper triangular.
Denote by $V\tilde \Lambda V^T$ the eigenvalue decomposition of $RTR^T$. 
Define $\tilde Q = \mathcal{Q}V$. 
Then, we have obtained $\hat D_{\tilde Q} = \tilde Q\tilde \Lambda \tilde Q^T$.
Moreover, $\tilde Q\tilde \Lambda \tilde Q^T$ is the $k$ truncated eigenvalue decomposition of $\hat D_{\tilde Q}$, since $\tilde \Lambda$ is diagonal and $\tilde Q$ is orthonormal as a product of orthonormal matrices. 
Finally, notice that $\hat D_{\tilde Q}$ is the reconstruction of $D$ using the approximated basis is $\tilde Q$.

\begin{figure}[htbp]
\begin{center}
\includegraphics[width=0.99\columnwidth]{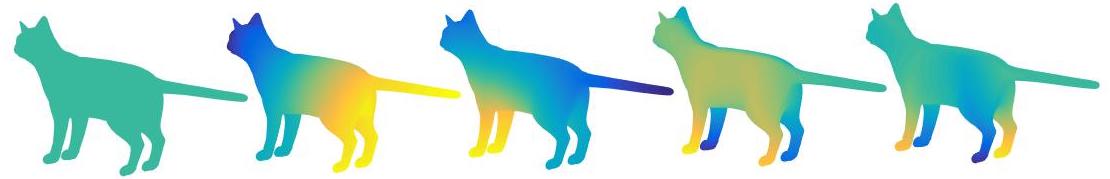}
\includegraphics[width=0.99\columnwidth]{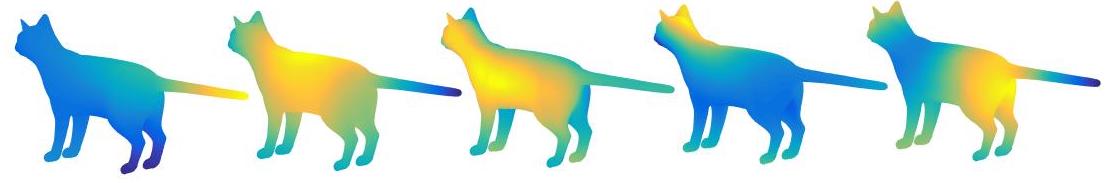}
\end{center}
\caption{The first $10$ eigenvectors of the basis.}
\label{fig:show_basis}
\end{figure}

Next, we measure how well $\tilde Q$ approximates the geodesic distances basis $Q$ by comparing the reconstructions $\hat D_{\tilde Q}$ and $\hat D_Q$.
Figure \ref{fig:compare_basis} demonstrates that $\tilde Q$ can be used instead of the optimal basis $Q$ with almost no effect on $\hat D_Q$.
Note that the above procedure supplies us the approximated basis $\tilde Q$ and the coefficients $\tilde \Lambda$ without the need to compute or store the entire matrix $D$, but only up to $p = 2k$ columns of it for computing $S$ and $T$ with the method from \cite{shamai2015classical}.
The geodesic basis vectors are no more than linear combinations of geodesic distances computed on the surface.
In Figure \ref{fig:show_basis} we vizualize the first $10$ basis vectors (columns of $\tilde Q$) on the Cat shape from TOSCA dataset.

\section{Geodesic descriptors}
\label{sec:GDD}
%
Denote by the diagonal matrix $W$ the square root of the diagonal matrix $\Lambda$, such that
\begin{equation}
 W_{ii} = \sqrt{\Lambda_{ii}}.
\end{equation}
Define 
\begin{equation}
X = QW, 
\end{equation}
 such that, 
\begin{equation}
D = XX^T.
\end{equation}
We term $X$ as the \textit{Geodesic Distance Descriptor} (GDD).
Note, that in general $X$ is complex.
$X$ can be used as a point descriptor that encodes the geodesic distances.
Since it stems from geodesic distances, it is not affected by isometric deformations of the shape.
$X$ holds all information of the geodesic distances $D$, and therefore could be used instead of
 $D$ to find the solution for \ref{eq:GH_inf}.
Moreover, $X$ is already represented in the basis $Q$ since $QQ^TX = X$,
 so only a few columns of $X$ contain almost all information encapsulated in $D$. 
This was demonstrated in Figure \ref{fig:compare_basis} by the reconstruction error of $D$
 as a function of number of eigenvectors.
The GDD can be therefore used for dimensionality reduction and simplification tasks that involve geodesic distances. 
In a sense, the GDD can be thought of a canonical form obtained using Multidimensional scaling \cite{borg2005modern}, 
 but without the embedding errors that occur because of flattening a curved surface. 

\begin{figure}[htbp]
\begin{center}
\includegraphics[width=0.49\columnwidth]{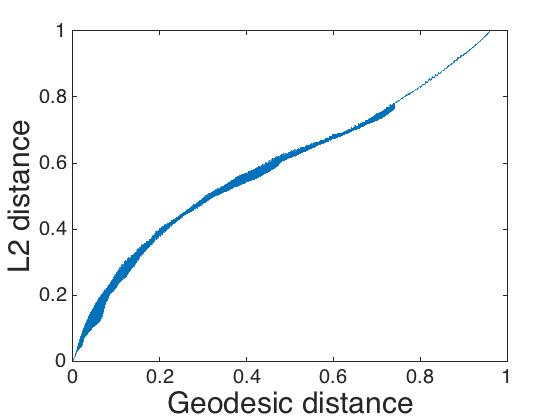}
\includegraphics[width=0.49\columnwidth]{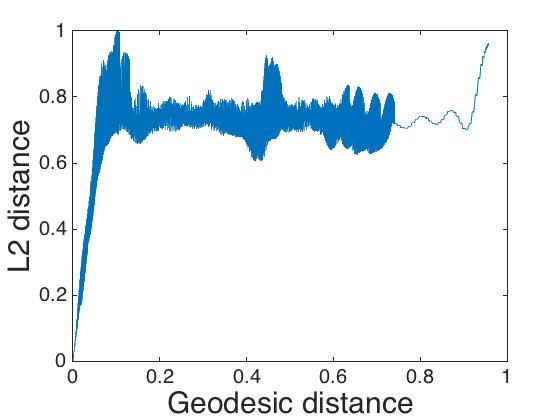}
\end{center}
\caption{$E_{ij}$ w.r.t $D_{ij}$. The left figure corresponds to $X$, and the right one corresponds to $\Phi$.}
\label{fig:descriptor_analysis_1}
\end{figure}

\begin{figure}[htbp]
\begin{center}
\includegraphics[width=0.49\columnwidth]{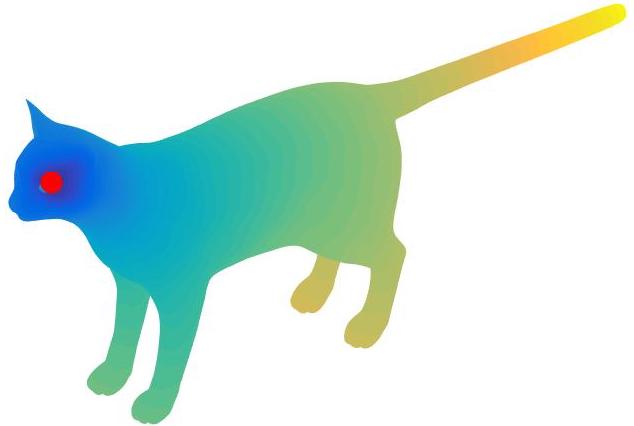}
\includegraphics[width=0.49\columnwidth]{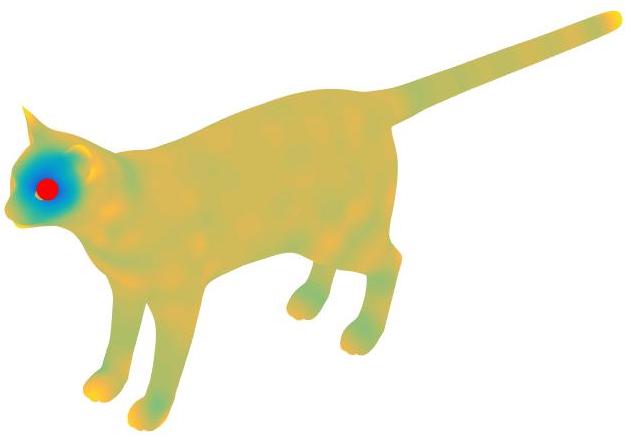}
\end{center}
\caption{ 
          The cats display the function $E_{i}(j)$ derived from $X$ (left) and $\Phi$ (right), 
          for a selected point $i$ marked in red.}
\label{fig:descriptor_analysis_2}
\end{figure}

Each point $i$ in the shape corresponds to a row $x^i$ in $X$. 
$x^i$ can be referred to as the descriptor of point $i$.
The descriptor of a point $i$ is invariant to vertex ordering of rest of the points. 
Hence, the GDD can be used as a point descriptor to find the correspondence between two shapes.
Denote by $E_{ij} = \|x^i - x^j\|_2$ the Euclidean distance between the descriptors of points $i$ and $j$. Figure \ref{fig:descriptor_analysis_1} shows the relation between $E_{ij}$ and the geodesic distance $D_{ij}$. 
For comparison, we show the same analysis for $\Phi$ instead of $X$, where $\Phi$ holds the eigenvectors of the LBO.
The upper cat shapes in Figure \ref{fig:descriptor_analysis_2} visualize $E_{i}(j)$ for a selected vertex $i$ marked in red, for both $X$ and $\Phi$.
It seems that the geodesic descriptors have monotonic relation with the geodesic distances, and thus more robust to large correspondence errors.

\subsection*{An alternative formulation}
Note that the eigenvalue decomposition has ambiguities.
When assuming no repeating eigenvalues, the eigenvectors in an eigenvalue decomposition are unique up to sign flips.
This can be formulated as $Q = Q_0C$, where $C$ is a diagonal sign matrix, and $Q_0$ and $Q$ are two possible eigenvectors matrices.
More generally, $C$ is an orthogonal matrix, representing rotation ambiguities that correspond to repeating eigenvalues.
In any case, the non-diagonal elements of $C$ correspond to the locations of the repeating eigenvalues on the diagonal of $W$.
It is therefore possible to swap between $C$ and $W$, such that 
\begin{equation}
X = QW = Q_0CW = Q_0WC = X_0C,
\end{equation}
where $X_0$ and $X$ are two possible derived GGD-s.
Therefore, the proposed geodesic distance descriptor is invariant to isometric deformations up to a rotation and reflection ambiguity $XC$.

Consider two isometric shapes with corresponding descriptors $X_1$ and $X_2$, and some correspondence encoded by the permutation matrix $P$. 
Plugging $D_1 = X_1X_1^T$ and $D_2 = X_2X_2^T$ in \ref{eq:GH_inf}, we obtain
\begin{equation}
\argmin_{P \in \pi(n)} \|PX_1X_1^TP^T-X_2X_2^T\|_\infty.
\end{equation}
It appears that this minimization can be reduced to solving
\begin{equation}
\label{eq:simp}
\argmin_{P \in \pi(n), C \in {\cal{U}}(n)}  \|PX_1C-X_2\|_{2,\infty},
\end{equation}
where $\|F\|_{2,\infty}$ stands for the maximal $L_2$ norm of any row in $F$, and ${\cal{U}}(n)$ is the set of $n \times n$ unitary matrices.
For isometric shapes, the two minimizations are equivalent. 
For nearly isometric shapes, the solution of \ref{eq:simp} approximates the the one of \ref{eq:GH_inf}, up to a some bound. 
If we manage to find a good solution to \ref{eq:simp}, it guaranties some bound on the minimizer of \ref{eq:GH_inf}.
More details regarding the bounds can be found in the supplamentary material.
In other words, solving the complex GH related minimization in \ref{eq:GH_inf} is nothing but matching the geodesic descriptors of two shapes, under the best rotation.

The minimization in \ref{eq:simp} can be efficiently solved using methods like Iterative Closest Point (ICP), with a quasi-linear complexity in the number of points, using efficient approximations such as kd-tree. 
Moreover, since $X_1$ and $X_2$ are already represented in the geodesic distance basis, it is enough to consider only their few $k$ first columns, with almost no effect on the solution.
In our experiments, $k = 50$ were enough for this task.
Note that ICP finds for each vertex in one shape a matching vertex in the other shape, rather then a bijective map. However, this makes more sense when dealing with two discrete shapes that might have been sampled differently from their corresponding manifolds.

\section{Initializations and post processing}
\label{sec:FGMDS}

Suppose we treat the descriptors $X_1$ and $X_2$ as two point clouds, where each row is a point, and the orthogonal matrix $C$ is a rotation and reflection of the points.
ICP (\cite{besl1992method}, \cite{chen1992object}) is an efficient optimization that attempts to compute the best match between two point clouds, under any rotation and reflection. 
In practice, ICP iterates between point match and rotation alignment:
\begin{enumerate}
\item For each point in the first could, find its nearest point in the second cloud, in terms of Euclidean distance.
\item Find the rotation a reflection that best aligns the matched points. This step has a closed form by using the Procrustes Theorem \cite{gower2004procrustes}.
\end{enumerate}
We suggest to find the solution of \ref{eq:simp} by applying ICP to $X_1$ and $X_2$. 
Note that indeed $C$ is defined as a unitary matrix and not orthogonal, and that $X_1, X_2$ are complex. Nevertheless, ICP could still be applied in the same manner, and $C$ could still be thought of a rotation and reflection matrix in the dual real and imaginary space. An alternative perspective would be to treat $X$ as a concatenation of its complex and imaginary parts. This would lead to an equivalent solution involving only real matrices.

ICP usually requires a good initialization. 
We propose the following alternatives.
\subsection*{Initialization using correspondence}
Assume we have some initial correspondence given as an output from another shape matching procedure. 
Plugging this initialization, it is possible to start with step $2$ of ICP, and continue to iterate. In our experiments, we used correspondences found by other methods as initializations, and managed to outperform any state of the art method.
\subsection*{Initialization using descriptors}
An alternative way is to start with some initial estimate of $C$, denoted here as $C_0$.
In functional maps, ICP between the eigenfunctions of the LBO was performed as a post processing step, while computing the initialization $C_0$ was the core of the method. 
First, different descriptors were computed for each of the shapes.
Then, $C_0$ was treated as a linear map between the coefficients of the descriptors in the LBO basis, and obtained using a simple least-squares minimization. 
At first sight, translating these steps to our problem is direct - 
compute $C_0$ using the coefficient of the descriptors in the basis of geodesic distances $Q$ instead of the LBO basis, and proceed similarly.
However, unlike functional maps, in our case the matrix $C$ encodes the deformation between the descriptors $X_1$ and $X_2$ which are not orthonormal bases (the columns are orthogonal but not normalized).
To that end, suppose that $f_1$ and $f_2$ are column vectors corresponding to some corresponding point descriptors on shapes ${\SSS}_1$ and ${\SSS}_2$. The orthogonal permutation matrix $P$ encodes a mapping between the shapes and can be therefore used for denoting $f_2 = Pf_1$. Since $P$ is orthogonal we can write instead $P^Tf_2 = f_1$ or $f_2^TP = f_1^T$.
Denote by $F_1$ and $F_2$ the coefficients of $f_1$ and $f_2$ in the geodesic distance basis representation, i.e, $F_1 = Q_1^Tf_1$ and $F_2 = Q_2^Tf_2$.
Suppose that we seek for some matrix $C$ that encodes the relation between $PX_1$ and $X_2$ as 
\begin{equation}
PX_1C = X_2.
\end{equation}
By multiplying both sides of the equation by $f_2^T$ we obtain
\begin{equation}
\label{eq:stam}
f_2^TPX_1C = f_2^TX_2.
\end{equation}
Notice that 
\begin{equation}
f_2^TPX_1 = f_1^TX_1 = f_1^TQ_1W_1 = F_1^TW_1.
\end{equation}
Hence, Equation \ref{eq:stam} can be further reduced to 
\begin{equation}
F_1^TW_1C = F_2^TW_2.
\end{equation}
Then, we could instead search for $C$ that encodes the relation between $F_1^TW_1$ and $F_2^TW_2$, which is independent of the mapping $P$. 
This can be defined as the minimization
\begin{equation}
\argmin_C\|F_1^TW_1C - F_2^TW_2\|.
\end{equation}
In other words, we propose to find an approximation to $C$ by repeating the procedure of functional maps, while using the basis $Q$ instead of the LBO basis, and multiplying the coefficients of the descriptors by the square-root of the eigenvalues, $W$.
\subsection*{Initialization using feature points}
Assume we have an initial set of $m$ points in shape ${\SSS}_1$ that correspond to $m$ points in shape ${\SSS}_2$.
To find the initial $m$ point correspondence, for example, in SGMDS it was suggested to first find in each shape a set of points that are locally farthest from the rest of the points, and then match the candidates using descriptors such as WKS \cite{aubry2011wave}. 
It was noted that $m = 5$ points were enough for a good initialization of SGMDS.

$m$ point correspondences can be considered as $m$ rows in $X_1$ that correspond to $m$ rows in $X_2$.
Denote the sub-matrices that correspond to these rows by $\hat X_1$ and $\hat X_2$.
Assume that the columns of $X_1$ and $X_2$ are ordered by the size of their corresponding eigenvalues, in a descending order.

For isometric shapes, assuming non repeating eigenvalues, $C$ would be a diagonal matrix. If the shapes are approximately isometric,
$C$ would have a sparse and diagonally dominant structure. This effect was already demonstrated in functional maps. 
Hence, it is possible to estimate $C$ by solving
\begin{equation}
\argmin_{C \in {\cal{U}}(n)} \|\hat X_1C - \hat X_2\|,
\end{equation}
while adding some off-diagonal penalty.
In fact, it is enough to estimate only the first rows and columns of $C$, and then obtain an initial correspondence using only the first corresponding columns of $X_1$ and $X_2$.
In our experimental setting, we used the same $m = 5$ point correspondences that were used in SGMDS, with which we approximated the $20 \times 20$ first rows and columns of $C$.

\subsection*{Post Processing}
The correspondence obtained using minimization \ref{eq:simp} is robust to large geodesic distances errors, 
 since they would penalize the objective function.
This was demonstrated earlier by showing that the GDD has a point signature that is unique to the point (Figures \ref{fig:descriptor_analysis_1} and \ref{fig:descriptor_analysis_2}).
Other bases or descriptors, however, could produce a signature with a more local nature.
These descriptors can further improve the solution be combining them with the GDD.
One simple way used in our experimental results is to refine the correspondence be performing ICP on the LBO basis, initialized with the correspondence obtained by our method.
As this basis appears to be better localizer than the GDD, superior results are obtained for correspondence when combining the two.
However, note that while the correspondence improves, this post-processing harms the approximation of the
 minimizer of \ref{sec:GH} (see experimental results). 
This is not surprising as the GDD is related to the minimization of \ref{sec:GH}.

\section{Results}
\label{sec:Results}
Throughout this section, we refer to our proposed method as the Geodesic Distance Descriptor (GDD), and compare it to methods discussed in the introduction and throughout the paper. 
In our experiments we used shapes from the publicly available datasets TOSCA \cite{bronstein2008numerical} and SCAPE \cite{dragomir2005correlated} that contain real and synthetic human and animal poses.
For accuracy comparison of shape correspondence we use the evaluation procedure proposed by Kim et al. \cite{kim2011blended}.
To the best of our knowledge, the state of the art methods for efficiently computing correspondences of nearly isometric shapes are Spectral-GMDS \cite{Aflalo2016}, functional maps \cite{ovsjanikov2012functional}, and Spectral-GF with ICSKM refinement \cite{shtern2015spectral}.
For GDD, we used $k = 50$ basis functions computed from $100$ samples in each surface. 

\begin{figure}[htbp]
\begin{center}
\includegraphics[width=0.45\columnwidth]{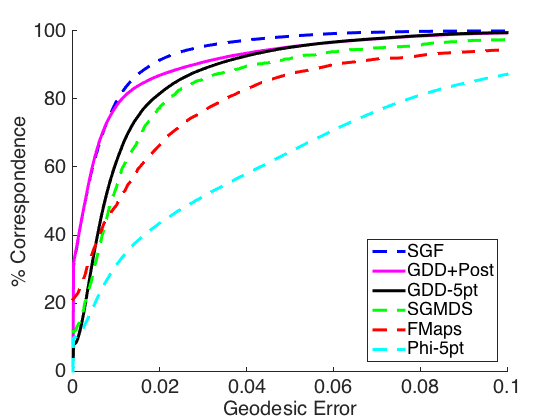}
\includegraphics[width=0.45\columnwidth]{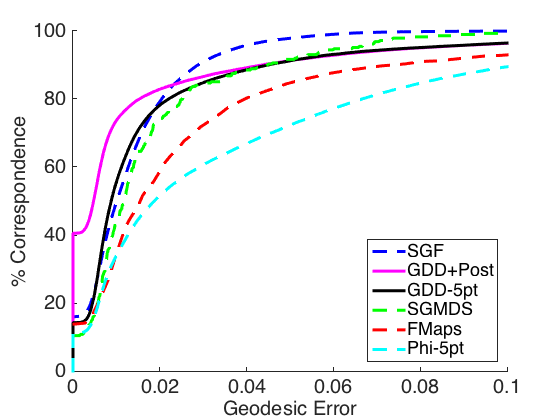}
\end{center}
\caption{Quantitative evaluation of shape correspondence methods applied to the shapes from the TOSCA and SCAPE datasets, using the protocol from Kim et al.}
\label{fig:kim-5pt}
\end{figure}

In our first experiment, given two shapes, we computed their point correspondence by applying ICP to their geodesic descriptors, using the same $5$ points initializations used for SGMDS. This was done with and without the post-processing step suggested in section \ref{sec:FGMDS} (GDD+post and GDD-5pt).
We compared the results to functional maps (FMaps), SGMDS and Spectral-GF (SGF).
We repeat the same experiment using the LBO basis instead of the $X$ (Phi-5pt).
The results are shown in Figure \ref{fig:kim-5pt}.
It can be seen that GDD performs better than FMaps with a much simpler initialization, without the need of the descriptors.
Compared to SGMDS, It can be seen that GDD performs better for the same initialization and with a much simpler procedure, without the spectral formulation for the optimization of the GH related minimization. 
In addition, our procedure required computing much less geodesic distances than SGMDS (see Section \ref{sec:NMDS}).

In the benchmark of Kim et al. the correspondence between shapes is assumed to be provided. 
The geodesic distance of each point mapped by the method we evaluate from what is referred to as true location is computed. 
The distortion curves describe the relative number of points falling within a relative geodesic distance 
 from what is assumed to be their exact location. 
Notice that the given ``exact'' location is, in fact, a subjective measure.
The distortion curves contain an intrinsic ambiguity of up to about $25 \%$ as there is no exact 
 isometry between objects at different poses.
 
In the next experiment, 
we 
 compute the correspondence $P$ as 
\begin{equation}
\label{eq:GH_F}
 \min_P \|PD_1P^T - D_2\|_F^2.
\end{equation}
Since we cannot really compute all pairwise geodesic distances, we evaluated 
 the result by considering only $1000$ randomly sampled rows and columns of the matrix inside the above norm.
We state the results in Figure \ref{fig:GH_approx}.
\begin{figure}[htbp]
\begin{center}
{\footnotesize
\begin{tabular}{|c|c|c|c|c|c|c|}
\hline
 & {\color{red}GT} & {SGF} & {SGMDS} & {Fmaps} & {\textbf{GDD}} & {Post} \\
\hline
{Horses} & {\color{red}470.1} & {518.7} & {707.4} & {1418.2} & \textbf{397.3} & {483.5}\\
\hline
{Victorias} & {\color{red}146.2} & {149.4} & {202.1}  & {147.6} & \textbf{128.5} & {150.7}\\
\hline
{Cats} & {\color{red}160.4} & {178.9} & {178.4}  & {189.7} & \textbf{123.5} & {154.5}\\
\hline
{Wolfs} & {\color{red}9.32} & {9.35} & {9.04} & {9.42} & \textbf{7.84} & {9.36}\\
\hline
{Centaurs} & {\color{red}153.5} & {174} & {151.5} & {771.9} & \textbf{122.7} & {154.4}\\
\hline
{Davids} & {\color{red}58.6} & {58.6} & {66.2} & {62.4} & \textbf{50} & {58.5}\\
\hline
\end{tabular}
}
\end{center}
\caption{Comparing correspondences as minimizers of \ref{eq:GH_F}.}
\label{fig:GH_approx}
\end{figure}
GDD and Post stand for GDD-5pt and GDD+post were discussed in the previous experiment.
It can been seen that the proposed GDD performs best as a method for approximating the minimizer of \ref{eq:GH_F}. 
Note that the post processing step damages the approximation, since it involves the LBO basis for 
 the sake of localizing the correspondence.
Surprisingly, GDD approximates \ref{eq:GH_F} even better than the ``ground truth'' correspondence provided
 by Kim et al.
It implies that these two measures might not always align for non-isometric shapes.

\begin{figure}[htbp]
\begin{center}
\includegraphics[width=0.19\columnwidth]{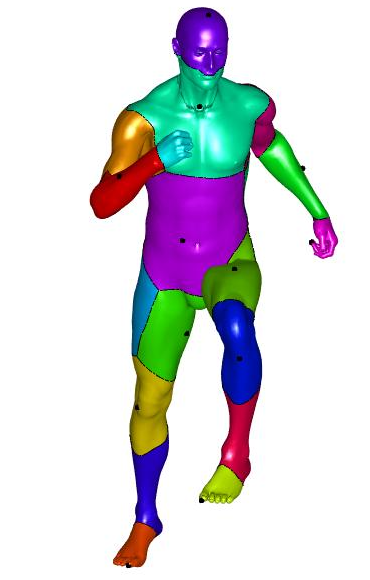}
\includegraphics[width=0.19\columnwidth]{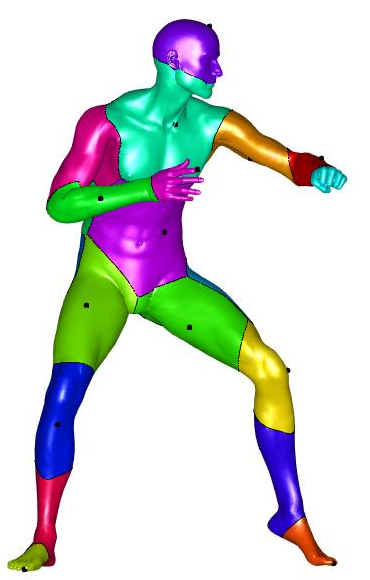}
\includegraphics[width=0.19\columnwidth]{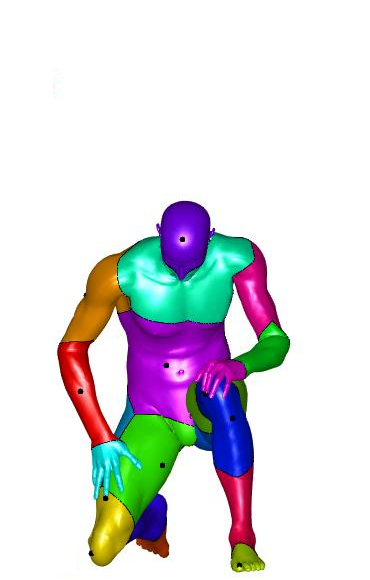}
\includegraphics[width=0.19\columnwidth]{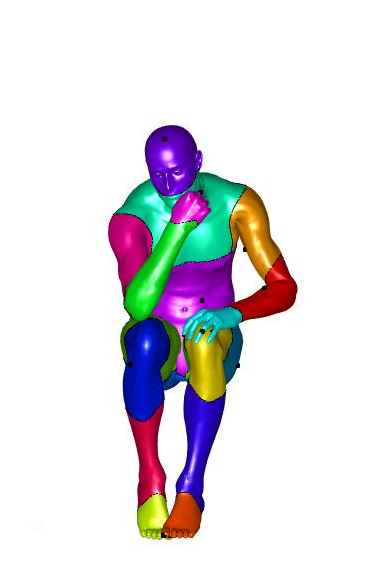}
\includegraphics[width=0.19\columnwidth]{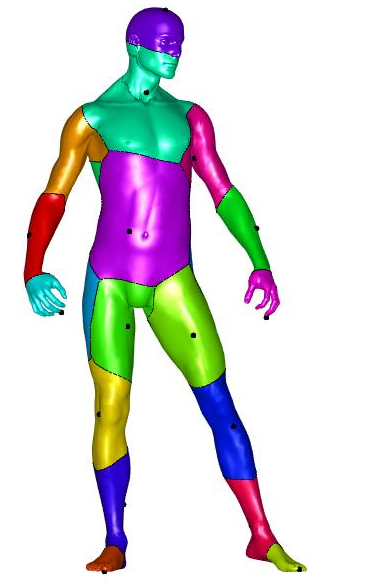}
\end{center}
\caption{}
\label{fig:visualize_corr}
\end{figure}

Next, we computed point-to-point correspondences between five Michael shapes from the TOSCA dataset. 
We then colored each shape according to the Voronoi regions of a set of $20$ points. 
Note that the Voronoi diagram was generated separately for each shape after mapping the set of $20$ points. 
The results are shown in Figure \ref{fig:visualize_corr}
Notice that some of the shapes were mapped into their intrinsic symmetries, as the objective \ref{eq:GH_inf} cannot
 differentiate between them and both solutions are optimal.

\begin{figure}[htbp]
\begin{center}
\includegraphics[width=0.45\columnwidth]{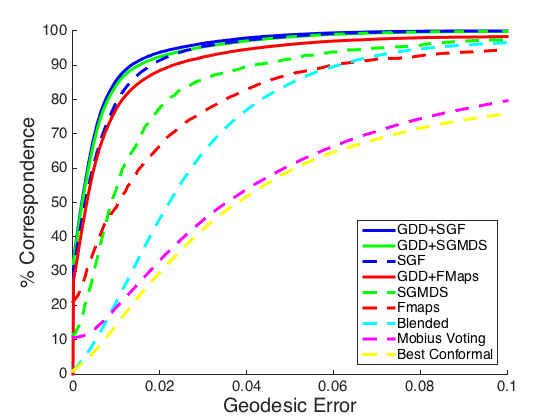}
\includegraphics[width=0.45\columnwidth]{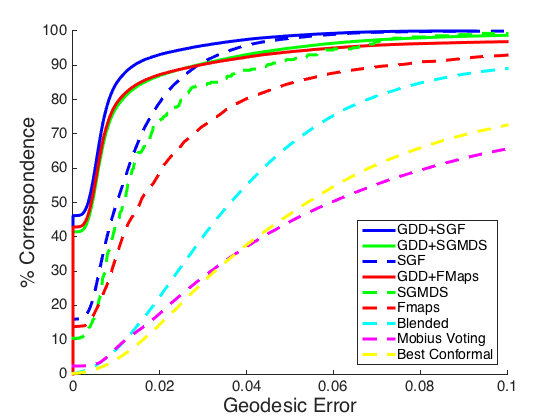}
\end{center}
\caption{Quantitative evaluation of shape correspondence methods applied to the shapes from
 the TOSCA and SCAPE datasets, using the protocol of Kim et al.}
\label{fig:kim-final}
\end{figure}

In our final experiment, we used the GDD to find the correspondence while initializing it with the
 correspondences computed by SGMDS (GDD+SGMDS), functional maps (GDD+FMaps) and Spectral-GF (GDD+SGF).
Figure \ref{fig:kim-final} shows the results.
It can be seen that using the GDD it is possible to successfully improve each of the methods,
 and thus reach state of the art results for the nearly isometric shape correspondence challenge.

\section{Conclusions}
The main contributions of this paper can be summarized by: 
\begin{enumerate}
\item Definition of a new basis that is optimized for geodesic distances representation. 
 	We also showed how to efficiently approximate it.
\item Definition of a generalized canonical form that does not suffer from embedding errors, 
	 and contains all information about the geodesic distances. 
 	We termed it as the geodesic distance descriptor.
\item An alternative formulation for the approximated GH distance related minimization of nearly isometric shapes
  using geodesic distance descriptors, that is both efficient and does not require relaxation or truncation of the 
  permutation matrices. 
\item Introduction of a shape correspondence procedure that obtains state of the art results for matching nearly isometric shapes.
\end{enumerate}

We introduced an efficient and accurate model for finding the best correspondence between two 
 metric spaces. 
The proposed method does not involve any relaxation or truncation of the eigenspace in which the permutation matrix is encoded.
The new formulation bridges the gaps between Spectral-GMDS, functional maps, and canonical forms, by introducing the 
 Geodesic Distance Descriptors.
The geodesic distance descriptor can be used for dimensionality reduction of tasks that involve geodesic distances. 
These distances are translated into a compact representation which is invariant to the order of vertices.
An optimal basis is proposed whose computation is based on 
  recent methods for geodesic distance approximations.
Experimental results show that while the accuracy of the metric space matching minimizer improves, the accuracy of the correspondence,
 as evaluated by a given manually-labeled pairs of corresponding points, does not necessarily improve. 
This finding suggests that the two measures are not necessarily the same.
Finally, it was shown that geodesic distance descriptor can be used to obtain state of the art matching results for
 nearly isometric shapes.

{\small
\bibliographystyle{ieee}
\bibliography{GDD}
}

\end{document}